# MACC: Cross-Layer Multi-Agent Congestion Control with Deep Reinforcement Learning

Jianing Bai*, Tianhao Zhang, Guangming Xie

Peking University

*Abstract*—Congestion Control (CC), as the core networking task to efficiently utilize network capacity, received great attention and widely used in various Internet communication applications such as 5G, Internet-of-Things, UAN, and more. Various CC algorithms have been proposed both on network and transport layers such as Active Queue Management (AQM) algorithm and Transmission Control Protocol (TCP) congestion control mechanism. But it is hard to model dynamic AQM/TCP system and cooperate two algorithms to obtain excellent performance under different communication scenarios. In this paper, we explore the performance of multi-agent reinforcement learning-based cross-layer congestion control algorithms and present cooperation performance of two agents, known as MACC (Multi-agent Congestion Control). We implement MACC in NS3. The simulation results show that our scheme outperforms other congestion control combination in terms of throughput and delay, etc. Not only does it proves that networking protocols based on multi-agent deep reinforcement learning is efficient for communication managing, but also verifies that networking area can be used as new playground for machine learning algorithms.

## I. INTRODUCTION

Various wired and wireless technologies have been widely used in communication system, such as Cellular network, 60 GHz mm Wave Wi-Fi [1], cognitive radio networks [2], [3], [4], Datacenter Networks [5] (DCN) and Internet of Things (IoT). Simultaneously, the widely used of new applications such as automatic driving, video streaming, on-line gaming and multiple network users contend over scarce communication resources creates higher Quality of Service (QoS) [6] requirements for the data transmission environment and poses new challenges on the design of network protocols in different layers. Consequently, there are reasons to research design of network protocols to do a better job of satisfy the new demands of emerging applications.

Among the user communication demands, one of the toughest problems is the mismatch between cross-layer protocols, which leads to network congestion at the bottleneck link. When there are too much packets flow into networks, to improve throughput, an intuitive approach is to increase the buffer enqueue rate of routers. However, rashly increasing buffer enqueue rate will increase the end-to-end delay and owing to the greedy nature of the TCP congestion control algorithm, it tends to increase the congestion window (CWND) and threshold to take up all the bandwidth until it detects packet loss. Then TCP congestion control scheme may retransmit more packet into networks, which lead to more severe congestion. So without communication and cooperation of different protocols, and just increasing sending rate and buffer enqueue rate will not benefit overall network transmission performance.

There are two main scheme that devoted to network congestion control. One is AQM, deployed at network layer, which is an effective policy to control the buffer-bloat problem. Another is TCP congestion control mechanism, deployed at transmission layer, which remains a really hot and active research topic in packet-switched networks. To achieve high performance of network communication, AQM should cooperate efficiently with TCP congestion control mechanism. However, it is hard to model dynamic AQM/TCP system and tune the parameters of both schemes to obtain good performance under complex and changeable network environment nowadays.

In the meanwhile, machine learning has made a huge break through and has been used in various application fields such as protein prediction, dialogue and automatic driven. Generally, machine learning can be divided into the following categories: supervised learning, unsupervised, semi-supervised learning and reinforcement learning. Supervised and unsupervised learning always train offline and need given datasets to obtain an optimal model. Semi-supervised learning is between supervised learning and supervised learning. It usually aims at the problem of missing sample labels or too few samples. While reinforcement learning can be trained online, which mapping action of agent to observed environment so as to maximize the function value of reward signal (reinforcement signal). Therefore, reinforcement learning has gained popularity for its applicability to real time world decision making. And a rich variety of networking protocols have been designed based on

bai@stu.pku.edu.cn

reinforcement learning algorithms, including routing protocol, TCP and AQM scheme based on RL. However, the need of of different layer protocol to learn control network congestion cooperation intelligently, propose challenge to AQM/TCP system design.

In this paper, we utilize our framework to design Multi-agent-based congestion control scheme (MACC). MACC is based on reinforcement learning [7] and two agents at different network layer to cooperatively generate a policy for mapping observed network statistics (e.g., latency, throughput) to decide packet sending rate, packet enqueue rate and drop rate. Our preliminary evaluation results suggest training MACC in simple, simulated environments is sufficient to generate congestion control policies that perform well also in very different network domains and which are comparable to, or outperform, recent state-of-the-art protocols.

In summary, we make following two contributions:

- We describe MACC, a deep RL based cross-layer congestion control scheme that automatically learns the effective strategies for one agent at transmission layer adjusting CWND and threshold, another agent at network layer adjusting enqueue rate and drop packet rate to cooperate achieve high throughput and low delay in an on-line manner. This fundamentally changes the design of previous protocols that no cross-layer cooperation and require fixed, manually selected rules.
- We propose a new kind of gym environment that can be used as multi-agent reinforcement learning algorithm playground. Therefore, the performance of various MARL algorithms can be tested in this environment.

The paper organized as follow: Section II gives a short overview of the related work. In Section III we present our system model of MACC. Then we give the experimental setup and elaborate the performance of the MACC in Section IV. Finally, the paper is concluded in Section V.

## II. RELATED WORKS

In this section, we first introduce some existing traditional network protocols and then discuss the work that exploits machine Learning for networking protocols, especially focus on reinforcement learning for congestion control mechanism.

**Rule-based protocols.** In the network layer, AQM is one of the intelligent network management methods in the network interface of a router or a switch [8]. In the traditional queue management methods, for example, RED [9] one of the early classic AQM schemes for congestion avoidance. CoDel [10], based on packet-sojourn time, tracks the minimum queueing delay experienced by the packets. Similar to CoDel, PIE [11] is also depending on the queueing delay, but is more robust with additional parameters. In the transfer layer, for example, Tahoe [12] and Reno [13], introduced three core phases in CC (i.e. slow start, congestion avoidance, and fast recovery), which become the foundation most TCP CC schemes build upon. Based on former works, there are many algorithms attempt to improve performance of congestion avoidance and recovery stages based on different signals. NewReno [14] close sender when all the lost packets are retransmitted and received confirmation. Selective ACK (SACK) algorithm [15] is based on selective ACKs and selective retransmission to Reno. Vegas [16], Fast active queue management scalable TCP (FAST) [17], low latencies TCP (LoLa) [18], and Timely [19], etc., treats increasing RTT as a congestion signal and adjust CWND to keep RTT in a desired range. Moreover, Veno [20], Africa [21], Compound [22], Libra [23] and Google Congestion Control (GCC) [24] combined loss and delay signals to evaluate congestion and most of them are based on Reno. To fully utilize multiple interfaces and balance the traffic load, multi-path TCP (MPTCP) also widely used in network communication. Accordingly, there are congestion control for MPTCP such as LAMPS [25] and AEPS [26].

**Learning-based protocols.** Over the past fifteen years, there many are efforts to implement intelligent solutions in congestion control to improvement the overall performance of network system. In the network layer, for example, RLGD [27], an AQM scheme controlling the queue length based on Reinforcement Learning Gradient-Descent. And QRED [28] adjust the maximum dropping probability according to the network situation based on RED scheme and Q-learning algorithm. RL-QDL [29] and RL-AQM [30] also present a new AQM algorithm based on RL to manage the network resources to keep the queueing delay low and the packet loss rate low at different communication situations. Apart from the above mechanisms adopted in routers, in the transfer layer, for example, Remy [31], based on the customized objective function consists of throughput and delay, attempts to find a mapping from a pre-computed lookup table, which cache all possible events to actions (including change in the cwnd). Instead of using hardwired mapping, PCC [32] and PCC-Vivace [33] leverages empirically observed performance metrics and online(convex) optimization in machine learning techniques to choose the best sending rate automatically. Contrary to clean-slate learning-based works that mentioned earlier, Orca [34] combines the traditional CC algorithm Cubic and DRL to computes the cwnd, which reduced the frequency of RL decisions and achieved suboptimal effects. Also, TCP-RL [35] uses RL and changes the initial congestion window and cwnd of TCP.

However, MACC fundamentally differs from these schemes as it attempts to use multi-agent RL techniques to help the cooperation of AQM and TCP congestion control and eventually boost the performance of the network communication performance.

## III. System Model

In this section we formulate cross-layer congestion control as a sequential decision-making problem under multi-agent reinforcement learning framework. Then we present an analytical description of the key components of MACC.

*A. Overview*

The framework of MACC is show in Fig.1. The multi-learning agent interacts with each other and the network environments based on deep RL algorithms. They are aimed at taking actions (e.g., varying enqueue rate and sending rate) based on environment states (e.g. transmission delay and throughput) and get higher reward, i.e., network communication performance, which received from environment feedback. Like any typical RL problem, MACC consists of the following elements:

- Agents: the first agent (i.e., agent[1]) is working at transmission layer and the second agent (i.e., agent[2]) is working at network layer.
- States: bounded histories of network statistics and defined as measurements that an agent can obtain from the outside environment. Here, the state is a unique profile of the network conditions evaluated through selected performance metrics (Part B).
- Actions: chosen by an agent at each time step, after perceiving its current state, according to a policy. In the context of congestion control, the action is the decision to change CWND, threshold, enqueue rate and drop packet rate (Part C).
- Reward: this reflects the desirability of the action picked. As we describe below, the reward is further specified by the value of a utility function. (Part D).
- Training algorithm: The purpose of the training algorithm is to learn the optimal policy to select certain action for each state. This is the central module of MACC as it is responsible for developing the congestion control strategies (Part E).

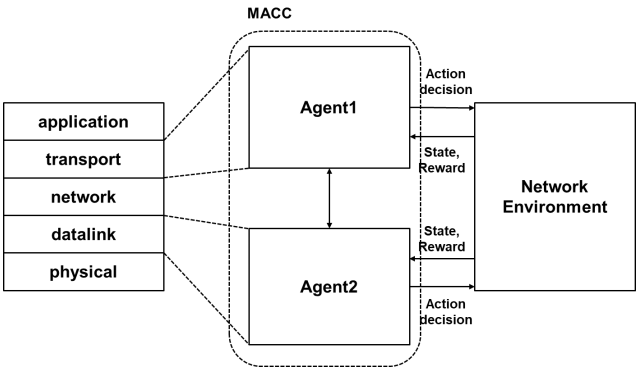

**Fig. 1.** Architecture of the proposed cross-layer multi-agent congestion control (MACC) schemes.

*B. States*

Networks greatly vary in terms of available bandwidth, latency, and loss rate, etc.

We consider three parameters to define the state for RL for agent[1]: (i) sum of segments acknowledged *sa*; (ii) bytes in flight *bif;* and (iii) round trip time of a packet *rtt*; agent[2]: (i) current queue length in number of packets *L*; (ii) dequeue rate $R_{deq}$; and (iii) queuing delay *d*. Time is divided into consecutive intervals. The state $s^i_t$ of each agent $i$ at time step $t$ is defined as:

$$v^1_t = \{sa_t, bif_t, rtt_t\},$$
$$v^2_t = \{L_t, R_{deq,t}, d_t\},$$

In our proposed MACC scheme, the agent's selection of the next actions is a function of a *fixed-length history* of the above statistics vectors. Considering a bounded-length history, instead of just the most recent statistics, allows our agent to detect trends and changes in network conditions and react more appropriately. Thus, the state at time *t*, $s_t$, is defined to be:

$$s_t = (v_{t-(k+d)}, \dots, v_{t-d}),$$

for a predetermined constant k > 0 and a small number d representing the delay between choosing a sending rate and gathering results.

*C. Actions*

In our formulation, the first agent is deployed at transmission layer, its actions translate to changes in sending rates. the sender can adjust its sending rate $x_t$, which then remains fixed throughout the step time intervals.

We choose to map our agent's output based on statistic vectors to change $cwnd_t$ and $threshold_t$ according to:

$$cwnd_t = \begin{cases} cwnd_{t-1} + segmentSize_{t-1}, & a_t \& 1 \\ cwnd_{t-1} + max(1, segmentSize^2_{t-1}), & else \end{cases}$$

$$threshold_t = \begin{cases} 2 \times segmentSize_{t-1}, & a_t < 3 \\ byteInFlightSum_{t-1}, & else \end{cases}$$

The second agent based on statistic vectors to decides a packet enqueue/drop rates $p\_drop_t$ and $p\_enqueue_t$ according to:

$$p\_drop_t = \begin{cases} 0.6, & a_t \& 1 \\ 0.4, & else \end{cases}$$

$$p\_enqueue_t = \begin{cases} 0.3, & a_t \& 1 \\ 0.7, & else \end{cases}$$

In order to find a better action in a certain state, we use explore/exploit strategy that allows that agent to either take an action based on its own selection (exploit), or sometimes take a random action chosen uniformly based on a certain probability (explore). For the explore/exploit strategy, we use ε-greedy algorithm starting from a high random action probability. In the first step of our network simulation, the exploring probability is set to 99%, which goes down to 0% through the episodes according to:

$$p\_exploring_t = 0.99^{memory\_counter}$$

for the integer variable $memory_counter \in [0, 2000]$ means the number of experience stored in replay memory. The agent takes an action in periodic interval $T_{int}$.

*D. Reward*

After taking an action, the RL agent waits for the next state $s_{t+1}$ during the interval $T_{int}$. The selected action is evaluated by a reward function. The reward resulting from chosen action of agents at a certain time may depend on the performance requirements of the specific application. Some applications (e.g., online gaming) might require very low latency while for others (e.g., large file transfers) high bandwidth is much more crucial.

We trained MACC with a linear reward function that rewards throughput while penalizing loss and latency. For agent i, we choose the following linear function as reward function $r^i$:

$$r^1 = segmentsAckedSum - bytesInFlightSum - cWnd$$
$$r^2 = dequeueRate - currQueueDelay - queueLength$$

Where *cWnd*, *dequeueRate*, *currQueueDelay* and *queueLength* is measured in packets per second, *latency* in seconds.

*E. Training*

We train our agent in an open-source gym environment described in detail in section 4. This environment simulates network links with a range of parameters. Our model was trained using the VDN [36] algorithm. And before model training the agent first detect if there exists trained model and reload model. Accordingly, after each iteration, agents save the trained model to corresponding path.

In our VDN model of MACC, agents store an experience $e_t = (s_t, a_t, r_t, s_{t+1})$ in tuple format to the replay memory at each time step $t$. Once the number of experiences in the replay memory exceeds the mini-batch size, the agent makes a random uniform selection of samples of the experiences from the memory. The agent minimizes the temporal difference error using mean square error loss function. For the Q-network, we employ the rectified linear unit activation function defined as $f(x) = max(0, x)$ at each hidden layer, and softmax function at the output layer to convert output layer to convert output into action probabilities.

*Algorithm 1* describes the training process of MACC.

**Algorithm 1** Proposed VDN-Based MACC Algorithm

**Require:**
 *gamma*: reward discount; *lr*: learning rate;
 *timestep*: the number of steps to run for per update;
 *episodes*: total number of samples to train on;
 *net_arch*: architecture of the neural network;

**Ensure:**

1: Ns3-gym environment initialization;

2: Set action space A;

3: Set state space S;

4: Reload or initialize network by *net_arch* for each agent

5: **for** e ∈ [1 : *episodes*] **do**

6:  Environment reset;

7:  **for** $t$ ∈ [1 : *timestep*] **do**

8:   **for** each agent $i$ ∈ [1 : *N*]**do**

9:    Execute action $a_{t,i}$ based on $\pi_{old,i}$ and $s_{t,i}$

10:    Observe next state $s_{t+1,i}$ and reward $r(s_{t,i}, a_{t,i})$

11:    Store the tuple $<s_{t,i}, a_{t,i}, r(s_{t,i}, a_{t,i})>$ into buffer

12:   **end for**

13:   $Q_t = \sum_1^N Q_t^i$

14:   Update $\pi_{old,i}$ by $Q_t$ for each agent

15:  **end for**

16: Save the trained model

17: Plot performance graph

## IV. EVALUATION

In this section, we first give the experimental setup and then elaborate performance of the MACC and make comparations with other well-known schemes. Finally, we give a brief analyze of the simulation results.

*A. Experimental Setup*

In [37] two well-known systems namely the ns-3 simulator and OpenAI Gym were combined to produce a benchmarking system called ns3-Gym. It simplifies feeding the RL with the data from the network in the ns-3 simulator and we also use this environment in this paper.

For the network topology we used in our experiments is indicated in Fig.2. That is on a network having some bottleneck links that result in packet losses when buffers are overloaded. The simple topology represents a real and complex network environment, where multiple flows sending by various devices compete for the battle-neck link's bandwidth. The congestion control protocol must dynamically adjust rending rate and buffer queue length to enable them sharing the network resources efficiently. Therefore, we deploy MACC scheme on each node, including senders, routers and receivers.

Simulation parameters used in the implementation are summarized in Table I.

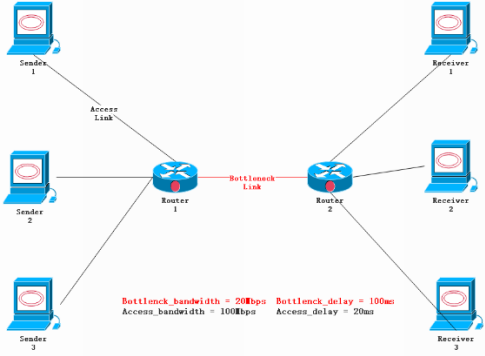

**Fig. 2.** The dumbbell-shaped network topology with N senders, N receivers and two routers simulation scenarios implemented in the ns-3 simulator. We set the bottleneck bandwidth as 20Mbps, delay as 100ms and packet error rate as 3%.

TABLE I. SIMULATION PARAMETERS

| Symbol | Parameter Name | Parameter Value |
|---|---|---|
| γ | Discount factor | 0.9 |
| α, μ | Initial learning rate | 0.95 |
| ε | Initial exploration rate | 0.1 |
| β | Weighing parameter | 0.8 |
| τ | Epoch duration | 0.2 seconds |
| - | Simulation time | 2000 seconds |
| - | Queue size | 385 packets |

### B. Compared Algorithms

In the simulation, we choose two well-known rule-based protocol and one learning-based protocol at network layer and transmission layer respectively to make a comprehensive comparison with our proposed MACC scheme. We note there are many alternative congestion control variants. However, we choose those protocols, shown in Table II, in our study because they both not only have large-scale real world deployment, but also represent the state-of-art solution in each category——

RED: the default algorithm to realize router congestion control, which mark the data packets and drops packets with a certain probability that arrive at the router.

CoDel: a packet-sojourn time based AQM, which tracks the minimum queueing delay experienced by the packets.

RL-QUE: the first learning-based AQM scheme, which uses DQN as its RL algorithm.

NewReno: classical and default congestion control protocol in use today and default CC algorithm in ns-3 platform as well, which is based on Reno.

BBR: employs two parameters, namely RTprop and BtlBw, to model the end-to-end network capacity.

RL-TCP: the classical DQN-based congestion control protocol.

TABLE II. COMPARED ALGORITHMS

| Network Layer Protocol | Transmission Layer Protocol |
|---|---|
| RED | NewReno |
| CoDel | BBR |
| RL-AQM | RL-TCP |

### C. Performance Analysis

Using the training algorithm mentioned in Section III, we trained universal neural networks for each agent and evaluated it in an online manner. We consider several critical parameters as our performance metrics including throughput, delay (RTT), stability of CWND and queue length. Throughput counts the amount of data successfully transmitted in a given unit of time, measured in Mbps. RTT measure packets transmission delay, which is the amount of time it takes for a signal to be sent plus the amount of time it takes for acknowledgement of that signal having been received. Stability of CWND and queue length calculates the mean absolute deviation (MAD) of CWND and queue in a given time interval according to:

$$\bar{d} = \frac{|x_1 - \bar{x}| + |x_2 - \bar{x}| + \ldots + |x_n - \bar{x}|}{n} = \frac{\sum_{i=1}^{n} |x_i - \bar{x}|}{n}.$$

The results are shown in Fig. 3-8. It is clear that after training a number of iterations, MACC is able to learn how to improve throughput and reduce delay shown in Fig.7 and Fig.8. And it tending to convergence after ten training sessions. We compared the tenth training performance of MACC with other protocols, it shown that MACC can achieve higher throughput in Fig.3 and lower delay in Fig.4. At the meantime MACC can keep its sending rate (i.e. CWND) and queue length at relatively stable level shown in Fig.7 and Fig.8.

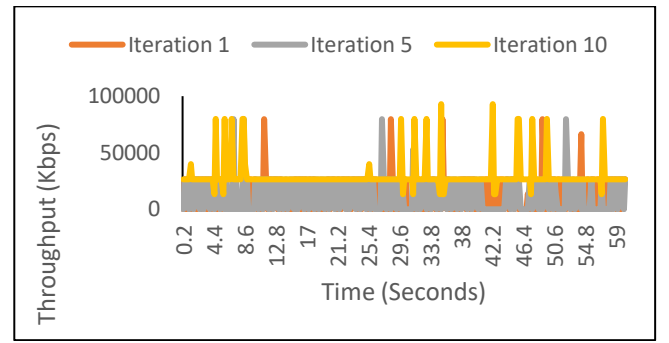

**Fig. 3.** Real-time throughput of MACC of three iterations.

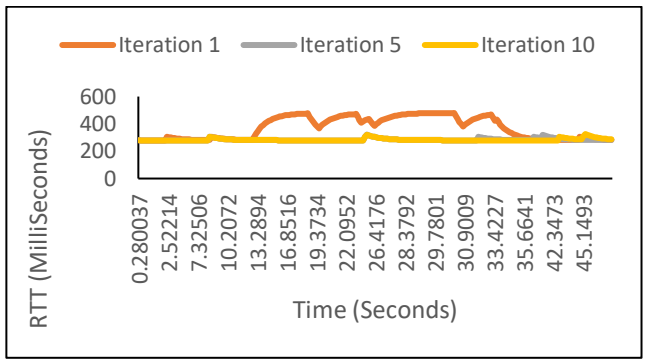

**Fig. 4.** Real-time RTT of MACC of three iterations.

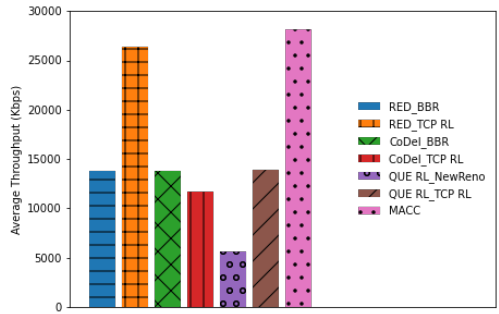

**Fig. 5.** Average throughput comparisons.

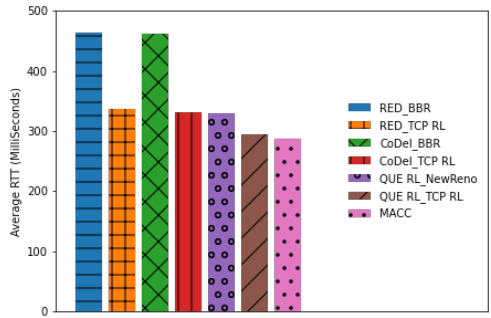

**Fig. 6.** Average RTT comparisons.

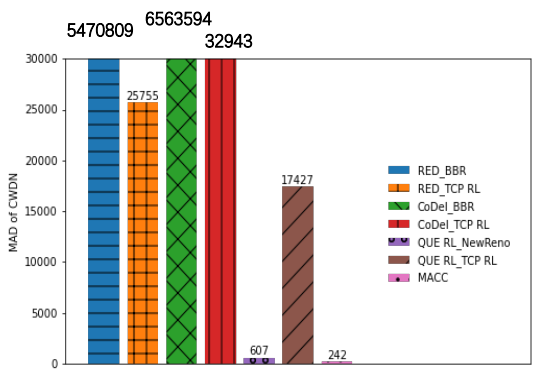

**Fig. 7.** Mean Absolute Deviation of CWND comparisons.

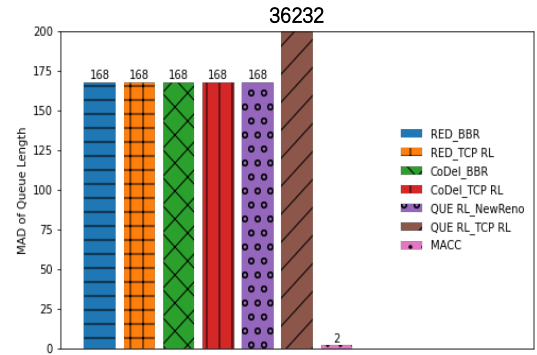

**Fig. 8.** Mean Absolute Deviation of queue length comparisons.

## V. CONCLUSION

We presented MACC, a cross-layer congestion control protocol powered by multi-agent deep RL to manage the network resources and keep low delay and high throughput at the same time. The proposed method has the advantage that it makes cooperation of different layer congestion control protocol and provides a new play-ground environment for multi-agent deep RL algorithm. The cross-layer congestion control is done by two Agent implementing a Deep Q-learning model. We evaluated the performance of our scheme by implementing it in various network scenarios in *ns-3* network simulator with *ns3-gym* package. And evaluated the performance of our scheme through simulation and compared with well-known combination schemes, RED, CoDel and RL-QUE with NewReno, BBR, and RL-TCP. We have shown that MACC can cooperate the different layer protocols and utilize the network resources efficiently.